# Spatially-multiplexed single-cavity dual-comb laser

**J. PUPEIKIS,[1,*] B. WILLENBERG,[1,*] S. L. CAMENZIND,[1] A. BENAYAD,[2] P. CAMY,[2] C. R. PHILLIPS,[1,*] AND U. KELLER[1]**

[1] *Department of Physics, Institute for Quantum Electronics, ETH Zurich, Auguste-Piccard-Hof 1, 8093 Zurich, Switzerland*
[2] *Centre de Recherche sur Les Ions, Les Matériaux et La Photonique (CIMAP), UMR 6252 CEA-CNRS-ENSICAEN, Université de Caen Normandie, 6 Boulevard Du Maréchal Juin, 14050, Caen Cedex 4, France*
*equal contributions
Corresponding author: pupeikis@phys.ethz.ch

**Single-cavity dual-comb lasers are a new class of ultrafast lasers which have a wide possible application space including pump-probe sampling, optical ranging, and gas absorption spectroscopy. However, to this date laser cavity multiplexing usually came to the trade-off in laser performance or relative timing noise suppression. We present a new method for multiplexing a single laser cavity to support a pair of noise-correlated modes. These modes share all intracavity components and take a near-common path, but do not overlap on any active elements. We implement the method with an 80-MHz laser delivering more than 2.4 Watts of average power per comb with sub-140 fs pulses. We reach sub-cycle relative timing jitter of 2.2 fs [20 Hz, 100 kHz]. With this new multiplexing technique, we could implement slow feedback on the repetition rate difference Δ$f_{rep}$, enabling this quantity to be drift-free, low-jitter, and adjustable – a key combination for practical applications that was lacking in prior single-cavity dual-comb systems. © 2022 Optica Publishing Group under the terms of the Optica Publishing Group Open Access Publishing Agreement.**

Dual optical frequency combs (or dual-combs for short) [1] were introduced soon after the optical frequency comb revolution [2–4]. In the time domain, dual combs can be understood as two coherent optical pulse trains with a slight offset in their repetition rates. Since their introduction, dual-comb sources and applications have been a major topic of research [5]. Dual-comb sources have many similarities to earlier laser systems used for pump-probe measurements. The idea of using two different repetition rates for sampling ultrafast phenomena was explored in the 1980s by the demonstration of the equivalent time sampling concept [6,7]. In this case, ultrafast dynamics are scaled down in the time domain to much slower equivalent time by the factor of $f_{rep}/\Delta f_{rep}$. Here $f_{rep}$ is the sampling rate and $\Delta f_{rep}$ the difference between the sampling and the excitation rate. This concept was soon implemented via a pair of mutually stabilized modelocked lasers, and is known as asynchronous optical sampling (ASOPS) [8]. A distinguishing feature between dual-combs and ASOPS laser systems is the relative timing and phase stability. The ASOPS lock performance is limited because after an electronic pulse train detection, only a lower order of the repetition rate is available. Conversely, the highest stability dual-comb performance is obtained after full dual optical frequency comb stabilization [2–4]. However, this is a complex and costly implementation, hindering wide adoption of dual-comb lasers.

The invention of dual-comb modelocking, where both combs are generated in a single laser cavity, is a promising solution to practical deployment of ASOPS and dual-comb systems. This novel modelocking approach was initially implemented in fiber [9] and solid-state [10,11] laser architectures. Because the pulses circulate inside the same cavity they experience similar disturbances, leading to correlated noise properties, which can be sufficient for practical applications [12]. Similarly, there is the potential for lower timing jitter compared to electronically locked ASOPS systems due to the common-path cavity architecture. Finally, ultra-low noise performance can be achieved when combined with a solid-state laser oscillator architecture due to the low intracavity nonlinearity, dispersion and losses supported by such oscillators [13,14].

To this date, single-cavity dual-comb laser operation was achieved as a trade-off in laser design. Example approaches include insertion of passive birefringent crystals into a cavity [10], misaligning a cavity with a birefringent gain element [15], splitting the laser gain bandwidth [16], or utilizing bidirectional operation of a ring cavity [9,11]. Lately, also spatially separated mode concepts which involve separate cavity end-mirrors were explored in thin-disk laser configurations [17,18]. However, in these latest implementations not all intracavity components are shared thus compromising common noise suppression.

The key parameter for deployment of single-cavity dual-comb lasers is the relative timing jitter. It directly relates to the phase-noise power spectral density of the repetition rate difference $\Delta f_{rep}$, and thus to variation in the comb line spacing. In addition, long-term drifts in $\Delta f_{rep}$ can be analyzed by measuring the time interval $1/\Delta f_{rep}$, with a frequency counter. While each individual measurement of this interval is subject to noise above the sampling rate of $\Delta f_{rep}$ [19], longer-term drifts are captured. For both types of measurements, a useful metric for system performance is the relative stability of the optical delay axis. For this, timing jitter can be normalized to the full optical delay range, while fluctuations in $\Delta f_{rep}$ can be normalized to the mean value of $\Delta f_{rep}$, thereby yielding a normalized quantity $\sigma(\Delta f_{rep})/\Delta f_{rep}$. Relative timing jitter measurements for single-cavity dual-comb lasers were only performed twice so far [17,19], yielding short-term optical delay axis stability of >5·10$^{-6}$ for the integration range [20 Hz, 100 kHz]. For long-term stability, reported values of $\sigma(\Delta f_{rep})/\Delta f_{rep}$ for single-cavity dual-comb lasers were in the >1·10$^{-5}$ range [16,20–23]. Such stability is comparable to electronically locked ASOPS systems [24].

In this letter we present a new approach to laser cavity multiplexing with which we could obtain 1.8·10$^{-7}$ [20 Hz, 100 kHz] short-term time axis stability (via relative timing jitter measurement) and 2.3·10$^{-7}$ $\sigma(\Delta f_{rep})/\Delta f_{rep}$ long-term stability over more than 5 hours (with a frequency counter). This translates to sub-cycle relative timing jitter and thus is a major step forward in the development of single-cavity dual-comb lasers.

We obtain spatial laser cavity multiplexing by inserting a monolithic device with two separate angles on the surface, such as a biprism. By installing the biprism at an appropriate position it is possible to adapt cavities that are optimal for a single comb operation into a dual comb. With this approach, we have obtained 2.4 W of average power per comb with less than 140-fs pulses at 80 MHz repetition rate from a single laser cavity. The relative timing noise between the pulses was characterized to be 2.2 fs [20 Hz, 100 kHz]. Furthermore, we have applied slow feedback on the multiplexing element to counteract low-frequency environmental disturbances and thus we could achieve a highly stable repetition rate difference stability with a standard deviation of 70 μHz on 300 Hz over more than 5 hours.

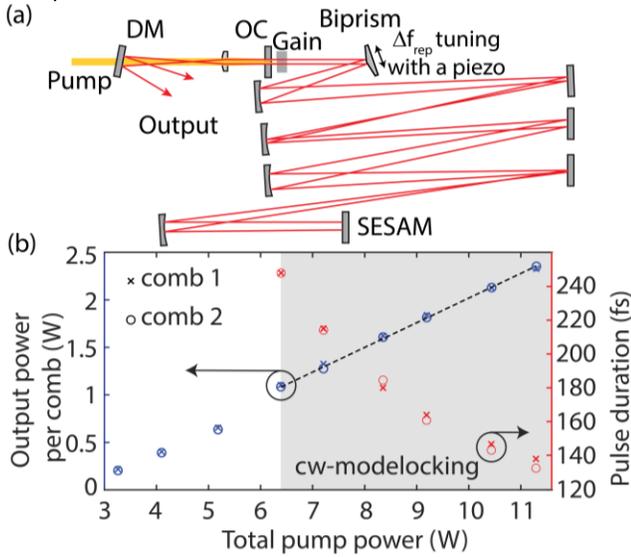

**Fig. 1.** (a) Cavity layout. DM: pump/laser dichroic, OC: 5.5% output coupler with high transmission for the pump (980 nm). The gain medium is an Yb:CaF$_2$ crystal with 4.5% doping [21]. SESAM parameters: $F_{sat}$=142 μJ/cm$^2$ and $\Delta R$=1.1%. (b) Laser output power and pulse duration evolution versus total pump power.

Our free-running dual-comb laser cavity layout is shown in Fig. 1(a). We use a multi-mode pump diode and an end-pumped cavity configuration, similar to polarization-multiplexed dual-comb lasers [21,25]. Here we obtain the dual-cavity condition by inserting a high reflectivity coated biprism with an apex angle of 179°. We obtain 1.6-mm mode separation on the gain (1/e$^2$ radii of 66 μm for the pump and 86 μm for the laser) and 1-mm separation on the SESAM (1/e$^2$ radius of 100 μm). Fig. 1(b) shows the individual comb performance as the pump power is scanned. The maximum operation point of this soliton-modelocked laser corresponds to 2.4 W average power in each output beam, with pulse durations of 138 fs (comb 1) and 132 fs (comb 2). The laser optical-to-optical efficiency is 40%.

We obtain self-starting modelocking of both combs. The laser output diagnostics at the highest output power are shown in Figs. 2(a-b) which indicate clean fundamental modelocking. The biprism's lateral position can be adjusted continuously over a short range with a piezoelectric actuator. The translation of the biprism allows tuning the repetition rate difference between the two combs from -450 Hz to 600 Hz with a negligible effect on the laser output performance (Fig. 2(c)). At larger excursions, mode clipping effects on the biprism apex led to a reduction in output power.

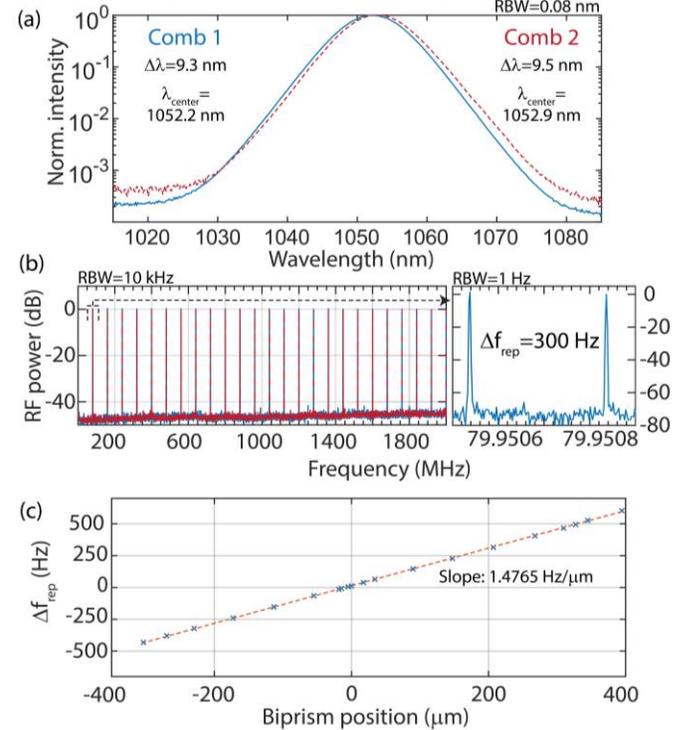

**Fig. 2.** (a) Laser output spectrum in log scale measured with an optical spectrum analyzer. (b) Normalized power spectral density of the photocurrent generated from a fast photodiode, analyzed with a microwave spectrum analyzer. Inset shows zoom on the first harmonic of the two radio-frequency combs. (c) The repetition rate difference versus the biprism lateral position using a picomotor (Newport).

Next, we evaluate the effectiveness of the common cavity approach for obtaining two pulse trains with low relative timing jitter. First, we perform the phase noise characterization to obtain the absolute timing jitter of each individual pulse train. We detect each pulse train on a fast photodiode (DSC30S, Discovery Semiconductors Inc.) and select the 6$^{th}$ repetition rate harmonic with a tunable band-pass filter. This signal is analyzed with a signal-source analyzer (SSA) (E5052B, Keysight). The obtained phase-noise power spectral density (PSD) and the integrated timing jitter are shown in Fig. 3. From the measurements we see that the phase noise PSD of both combs look virtually identical.

To measure how much the absolute timing jitter is correlated between the two pulse trains we have developed a relative timing jitter measurement technique based on comb line beating with two single-frequency cw lasers [19]. This relative timing jitter measurement technique reveals the uncorrelated noise of a free-running dual-comb laser at any repetition rate difference. The obtained uncorrelated relative timing jitter is plotted in Fig. 3 as a black line (see Supplement 1 for measurement details). We find the relative timing jitter spectrum to be more than 25 dB lower than the absolute timing jitter spectrum across almost all frequencies, suggesting excellent common timing noise suppression due to the single cavity architecture. The integrated relative timing jitter is found to be 2.2 fs [20 Hz, 100 kHz]. This demonstrates that sub-cycle relative timing jitter can be obtained from the free-running laser cavity even over relatively long acquisition times.

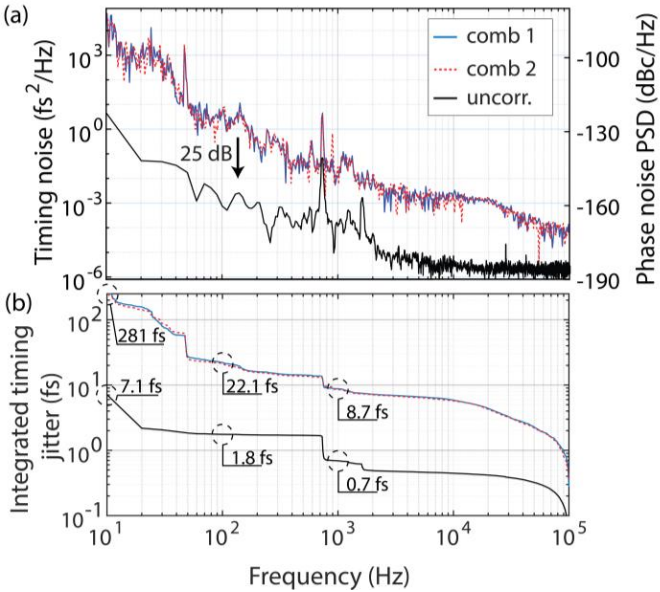

**Fig. 3.** (a) Absolute (red and blue) timing noise of each individual pulse train as measured using a signal-source analyzer. Relative timing jitter (black) between the two pulse trains as measured using the multiheterodyne measurement described in Supplement 1. (b) Timing jitter as acquired by integrating the timing noise curves.

A low relative timing jitter is one of the key prerequisites for coherent interferogram averaging in high-resolution dual-comb spectroscopy. However, the relative optical phase between the two pulse trains still can fluctuate and broaden the relative radio-frequency (RF) linewidth. We perform a relative optical phase stability analysis as detailed in the Supplement 1. We find that the RF comb line linewidth is 425 Hz at 1 kHz and broadens to 2.57 kHz at 1 Hz. This linewidth broadening occurs due to mechanical resonances at 700 Hz and 1600 Hz which are visible also in the relative timing jitter measurement since the laser is not mechanically isolated from the environment. Therefore, comb line resolved measurements would require optimized mechanical design, operation at a repetition rate difference $\Delta f_{rep}$ above 1 kHz or tracking of individual comb lines with a cw laser. In Supplement 1 we show how such a comb-line-resolved measurement can be performed using this laser. Due to the sub-cycle jitter and relative phase correction the interferograms can be coherently averaged even over a 1-second-long measurement time window.

We have developed this 80-MHz laser for pump-probe spectroscopy applications, where the high-peak power of the laser can be used to directly excite nonlinear processes. The 80-MHz repetition rate enables a large 12.5-ns delay scan range, and the ultra-low relative timing jitter can be used for precise time axis calibration. In our opinion much more suited configurations for single-cavity dual-comb spectroscopy involve higher repetition rates and repetition rate differences, such as [12,19], and thus exploring this multiplexing technique presented here in that regime will be a subject of the future work. In this publication we focus on deploying this laser for rapid pump-probe sampling applications.

One of the key parameters for any rapid-sampling application is the relative intensity noise (RIN) of the laser. We analyze the RIN of our laser in the following high-dynamic range measurement configuration. We use a photodiode which we expose with 10 mW of average power from each comb at a time. To acquire RIN spectra we perform baseband measurements with the SSA. First, we measure low frequency components (<200 kHz) with a low-noise transimpedance amplifier (DLPCA-200, Femto). To measure higher frequency components, we split the AC and DC parts of the signal with a bias-tee (BT45R, SHF Communication Technologies AG). The AC part is amplified with a low-noise voltage amplifier (DUPVA-1-70, Femto). The two measurements are stitched together to obtain a complete RIN spectrum for each comb as shown in Fig. 4. We find that the integrated RIN of each comb is < $3.1 \times 10^{-5}$ [1 Hz, 1 MHz].

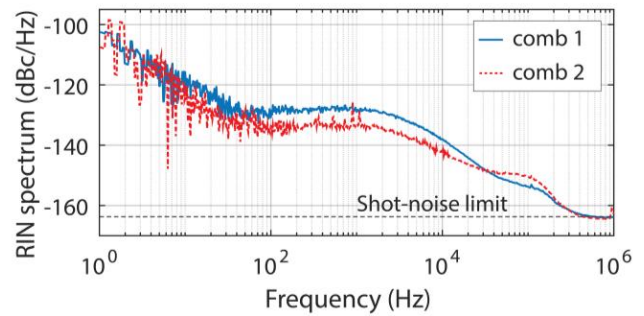

**Fig. 4.** Relative intensity noise spectrum of each comb. Shot-noise limit was calculated based on photodiode specifications and input power.

To enable pump-probe spectroscopy applications with the laser, we have coupled it with an optical parametric oscillator (OPO) for one of the output beams. The OPO enables wavelength versatile multi-color pump-probe measurements. Also, since the OPO is pumped synchronously, the relative timing between the two pulse trains remains preserved. We have designed an OPO using a PPLN crystal (HC Photonics) which is signal resonant at 1600 nm. By pumping with 2 W of comb 1 we could obtain 876 mW of signal radiation. At the same time, we have also generated the second harmonic of the OPO signal to obtain 800 nm radiation with a measured pulse duration of 151 fs and 390 mW of average power. Comb 2 from the oscillator can be easily frequency doubled to obtain 526-nm radiation, making this laser source an ideal nonlinear spectroscopy tool at various wavelengths.

To obtain long-term repetition rate difference stability despite any environmental changes we implemented a slow feedback loop. Part of the comb 1 and comb 2 power is sent to a BBO-based optical cross-correlator. We use a frequency counter to track the repetition rate difference fluctuations by counting the time between the cross-correlation signals. This measurement closely corresponds to the period jitter of the laser [19,21,25]. For this purpose, we use a field-programmable gate array (FPGA) module which can acquire the repetition rate difference with a precision <$10^{-6}$ at the rate of $\Delta f_{rep}$. We use a computer to control the piezoelectric actuator with 20-bit resolution leading to 5 pm theoretical step in 5 μm range.

To demonstrate the relative long-term stability of the two multi-color pulse trains we measure the repetition rate difference with another optical cross-correlation setup as shown in Fig. 5(a). We cross-correlate the frequency doubled OPO output (800 nm, comb 1) with the direct laser output at 1053 nm (comb 2). Over the longer than 5-hour time window we find that the repetition rate difference standard deviation was 70 μHz as shown in Fig. 5(b).

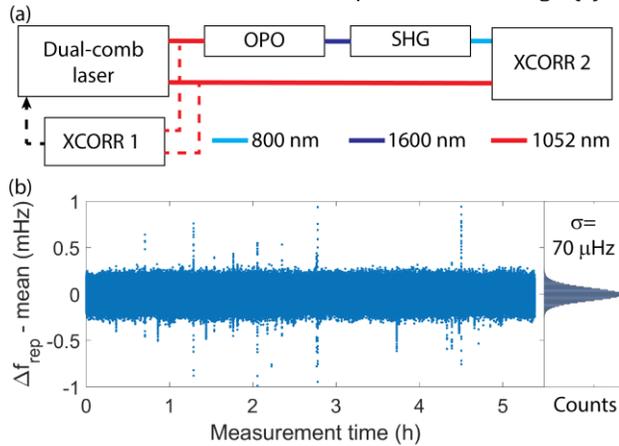

**Fig. 5.** (a) Multi-color equivalent time sampling setup with two optical cross-correlators (XCORR). XCORR 1 is used to provide slow feedback to the laser and XCORR 2 is used to perform an out-of-loop measurement. (b) Long-term repetition rate difference $\Delta f_{rep}$ stability with XCORR 2. The $\Delta f_{rep}$ was set to 300 Hz.

In summary, we have shown a novel laser cavity multiplexing method which allows for two spatially separated, yet quasi-common-path cavity modes within the same oscillator. We could achieve simultaneous modelocking with average powers above 2.4 W for each comb and pulse durations less than 140 fs. We have also characterized the relative timing jitter to be in the sub-cycle range for integration bandwidths of 20 Hz to 100 kHz. We further have coupled this powerful solid-state laser with an OPO to obtain a multi-color configuration for pump-probe sampling applications. To remove any slow environmental drifts which could change the repetition rate difference, we have implemented a slow cross-correlation-based feedback loop on the biprism position which allowed us to obtain excellent long-term performance of the dual-comb over several hours. Therefore, our system combines the benefits of both approaches: the high passive stability and simplicity of common-cavity dual-comb lasers, together with the immunity to drifts of locked laser systems. We obtain $1.8 \cdot 10^{-7}$ [20 Hz, 100 kHz] short-term time axis stability and $2.3 \cdot 10^{-7}$ $\sigma(\Delta f_{rep})/\Delta f_{rep}$ long-term stability over more than 5 hours, which is record-level performance for a single-cavity dual-comb laser. Our results demonstrate the usefulness of the new laser cavity multiplexing approach and show its great potential to revolutionize how equivalent time sampling and dual-comb systems are implemented.

**Funding.** Schweizerischer Nationalfonds zur Förderung der Wissenschaftlichen Forschung (40B1-0_203709, 40B2-0_180933); H2020 European Research Council (966718).

**Acknowledgments.** We thank FIRST clean room facility at ETH Zurich and Dr. Matthias Golling for the SESAM, and we acknowledge Alexander Nussbaum-Lapping for his characterization of the SESAM.

**Disclosures.** The authors declare no conflicts of interest.

**Data availability.** Data underlying the results presented in this paper are available in Ref. [26].